\RequirePackage[2020-02-02]{latexrelease}
\documentclass
[superscriptaddress,secnumarabic,amssymb,amsmath,nobibnotes,aps,prd,showkeys,showpacs,twocolumn,nofootinbib]{revtex4}%
\usepackage[T1]{fontenc}
\usepackage{graphicx}
\usepackage{hyperref}
\usepackage{amsmath}
\usepackage{amsthm}
\usepackage{amsfonts}
\usepackage{xcolor}
\usepackage{amssymb}
\usepackage{stmaryrd}
\usepackage{soul}

\usepackage[toc,page]{appendix}

\def\be{\begin{equation}}
\def\ee{\end{equation}}
\def\bea{\begin{eqnarray}}
\def\eea{\end{eqnarray}}

\begin{document}

\def\ra{\longrightarrow}
\def\ci{\mathcal I}
\def\ca{{\mathcal A}}
\def\cb{{\mathcal B}}
\def\cc{{\mathcal C}}
\def\cd{{\mathcal D}}
\def\ch{{\mathcal H}}%
\def\cm{{\mathcal M}}
\def\cv{{\mathcal V}}
\def\cw{{\mathcal W}}
\def\cR{{\mathcal R}}
\def\hr{{\hat R}}
\def\cH{{\mathcal H}}
\def\fH{{\mathfrak H}}
\def\cT{{\mathcal T}}
\def\da{{\dot{a}}}
\def\dph{{\dot{\Phi}}}
\def\aa{{\mathfrak a}}
\def\xx{{\mathfrak y}}
\def\tT{{\mathfrak t}}
 
\title{Scalar-tensor cosmologies in a minisuperspace\\ formulation: a case study}

%
 
\author{A. Borowiec and A. Kozak}
	\email{andrzej.borowiec@uwr.edu.pl, aleksander.kozak@uwr.edu.pl}
\affiliation{ University of Wroclaw, Institute of Theoretical Physics, pl. Maxa Borna 9, 50-206 Wroclaw, Poland\\
}


\begin{abstract}
We revisit a general minisuperspace (MSS) formalism for scalar tensor (ST) FLRW type cosmological models in arbitrary frame with perfect fluid source. We discuss how to impose Cauchy data on the corresponding dynamical system in order to reconstruct standard ($\Lambda$CDM) cosmological model.
So far, the integrability of such models has been extensively studied in the Einstein or Jordan frames mainly. We extend these studies  to arbitrary frame taking into account non-minimal coupling between matter and gravity. To this aim we explore a gauge freedom  associated with a choice of lapse function.
 We show that particular isothermal MSS coordinates are related with Einstein frames representing solution equivalent classes and have some invariant meaning. This provides new universal framework for investigating cosmic evolution in arbitrary frame: analytical solutions obtained in the Einstein frame can be transformed back to the physical frame by making use of a conformal transformation and field redefinition. We also show how such technique can be useful when applied to  Wheeler-DeWitt quantum cosmology.
  
\end{abstract}


\maketitle

\bibliographystyle{plain}
\maketitle
\section{Introduction}

Scalar-tensor theories (STT) introduce a non-minimally coupled, dynamical scalar field that, together with the metric tensor, mediates gravitational interaction \cite{so2}-\cite{borowiec2020}. They gained interest after certain shortcomings of general relativity had become manifest, such as the need for an inflationary period or the currently observed accelerated expansion of the Universe. Introduction of a scalar field that is non-minimally coupled to the curvature allows one to create scenarios in which the dynamics of the field, accounting for various phenomena, becomes an inherent feature of the theory of gravity itself \cite{sami2017}-\cite{bauer}. It was shown that inflationary behavior can be realized by addition of Higgs field \cite{syksyr}-\cite{bezrukov}, both in the metric and in Palatini approach, and the accelerated expansion is well described by extended quintessence models \cite{bartolo1999,bauer2011,wang}, in which the role of cosmological constant is played by a dynamical scalar field, whose energy density tracks density of radiation and then density of dark matter. 

In the most general case, the field can be coupled both to the curvature and to matter fields present in the theory. By means of Weyl (or conformal) transformation of the metric tensor one can always choose a parametrization, in which the scalar field is non-minimally coupled only to either the gravitational or matter part of the action functional. Different parametrizations are called 'conformal frames', and the two most commonly used in the literature are Einstein frame, with the field coupled to matter, and Jordan frame, where non-minimal coupling to curvature is present \cite{flan}, \cite{calmet2007}. Usually, Weyl transformations are used as a mathematical tool, allowing one to choose a frame in which the calculations are easier, and then transforming back to the frame considered to be physical. There is no agreement as to which frame is the right one, various arguments in favor of either can be found in the literature \cite{so9}-\cite{kamen2015}. Some authors, however, claim that all conformally-related frames are in fact different representations of the same physical theory, and all quantities that can be connected to observations should be expressed in terms of variables invariant under Weyl rescaling \cite{jarv}, \cite{burns2016}. 

A superspace formalism for general relativity was initially introduced for the purposes of quantizing gravity and in particular quantum cosmology. 
It originates from canonical quantization of gravity by means of functional Wheeler-DeWitt Schr\"{o}dinger type equation \cite{deWitt}. For Bianchi cosmological models however, it is possible to simplify some considerations by replacing  
infinite dimensional (functional) superspace with a finite dimensional minisuperspace. Thus MSS approach reduced the quantum cosmology to quantum mechanical systems of single particle with different potentials and kinetic energy terms as a toy model see e.g. \cite{Misner}-\cite{Barrow2015}.
Later on, the corresponding classical mechanical system in MSS formulation
proved to be extremely effective in studying (classical and quantum) integrability of scalar-tensor cosmological models 
in a specific (usually Einstein or Jordan) frame \cite{mss06}-\cite{Vernov2018}, and finding out their exact solutions,
see e.g. \cite{Fre}-\cite{Barrow2018} and references therein. Hamiltonian framework is also very useful in dynamical system analysis of such systems.
Our aim in the present paper is to extend these considerations to a broader frame-independent context.  

Conformal transformations affect the MSS formulation of theory, in particular in the case of FLRW metric and description of cosmic evolution. Weyl rescaling changes properties of geometry of the minisuperspace, as one can always find such a parametrization for which its curvature vanishes or the system possesses non-trivial first-order integrals of motion. In this paper, we will investigate the issue of dependence of MSS variables on the choice of conformal frame. The outline of the work is the following: first, we will write the most general action for STT and obtain equation of motion. Then, MSS formulation will be presented. To make the considerations as general as possible, we keep the lapse function arbitrary, which will allow us to obtain the constraint (Friedmann) equation. It will be shown that the lapse function can be thought of as a conformal factor for the MSS metric. Introducing such a metric and reparametrizing the potential will allow us to simplify the resulting equations of motion for the scalar field and the scale factor; for this purpose, making use of the constraint equation will be necessary. Also, universal coordinates will be introduced and used to construct an MSS Lagrangian manifestly invariant under Weyl rescaling of the space-time metric. At the end, we will focus on analysis of metric $f(R)$ and hybrid theories of gravity, and show that it is possible to find first-order integrals of motion by making a specific choice of the conformal variables.


\section{Effective minisuperspace description for scalar-tensor theories of gravity}

\subsection{STT \& FLRW cosmology with lapse function}
The most general action for scalar-tensor theories of gravity in the metric approach can be written as
(see e.g. \cite{borowiec2020}, \cite{jarv} for the same convention)\footnote{In fact, as shown recently \cite{borowiec2020}, such metric STT allow to describe as well Palatini and hybrid metric-Palatini ST  gravities.}:
\begin{equation}\label{action1}
\begin{split}
    S[g_{\mu\nu},\Phi,\chi] & = \frac{1}{2\kappa^2}\int_\Omega d^4x\sqrt{-g} \Big[\ca(\Phi)R -\cb(\Phi)g^{\mu\nu}\partial_\mu\Phi\partial_\nu\Phi \\
    &-\mathcal{V}(\Phi)\Big] + S_\text{matter}\left[e^{2\alpha(\Phi)}g_{\mu\nu}, \chi\right],
\end{split}
\end{equation}
where the given functions $\{\ca(\Phi), \cb(\Phi), \cv(\Phi), \alpha(\Phi)\}$ are frame parameters. Usually $\ca(\Phi)$ is assumed to be positive function as an effective gravitational constant,  $\alpha(\Phi)$ describes non-minimal coupling between the gravity and matter source. Varying the action (\ref{action1}) with respect to the metric and the scalar field one obtains general field equations for the theory
\begin{subequations}
	\begin{align}
		\begin{split}
			& \ca G_{\mu\nu} + \left(\frac{1}{2}\cb  + \ca''\right)g_{\mu\nu}g^{\alpha\beta}\partial_\alpha\Phi\partial_\beta\Phi -\left(\cb + \ca''\right)\partial_\mu\Phi\partial_\nu\Phi \\
			& - \ca'(g_{\mu\nu}\Box - \nabla_\mu\nabla_\nu)\Phi +\frac{1}{2}\cv g_{\mu\nu} = \kappa^2 T_{\mu\nu},
		\end{split} \\
		\begin{split}
			& 2[3(\mathcal{A}')^2 +2\mathcal{A}\mathcal{B} ]\Box\Phi + \frac{d [3(\mathcal{A}')^2 +2\mathcal{A}\mathcal{B} ]}{d\Phi} (\partial\Phi)^2+4\mathcal{A}'\cv\\
			&-2\mathcal{A}\mathcal{V}'=2\kappa^2\,T (\mathcal{A}'-2\alpha' \mathcal{A} )\,,
		\end{split}
	\end{align}
\end{subequations}
where $T_{\mu\nu}=-\frac{2}{\sqrt{-g}}\frac{\delta S_{\text{matter}}}{\delta g^{\mu\nu}}$
denotes the source matter stress-energy tensor and $\Box=g^{\mu\nu}\nabla_\mu \nabla_\nu$. Remarkably, their solutions can be transformed from one frame to another by making use of metric conformal transformation and scalar field redefinition. Thus the totality of all frames can be divided into solution equivalent classes \footnote{As discussed many times in the literature, see e.g. \cite{so9} - \cite{kamen2015}, solution or mathematically equivalent frames are not physically equivalent.}.

Here, we focus our attention on cosmological applications of the theory. To this end, we take the metric to be Friedmann-Lema\^{i}tre-Robertson-Walker (FLRW) metric with a reparametrized time component (lapse function), together with the assumptions of homogeneity and isotropy of the Universe, which means that the only dynamical entities entering the action, the scale factor and the scalar field, depend on the local (coordinate) time $t$. The reparametrization will allow us to assume gauges that will prove convenient when analyzing different physical situations.
\begin{equation}\label{metric:metric1}
g_{\mu\nu} = \text{diag}\left(-N(t)^2, \frac{a(t)^2}{1-Kr^2}, a(t)^2r^2, a(t)^2r^2\sin^2\theta\right)
\end{equation}
For this metric ($K\in \{-1,0,1\}$ denotes spatial curvature corresponding to an open, flat or closed universe), the Ricci scalar reads as ($\dot{( )}=\frac{d}{dt}$ ):
\begin{equation}
R = \frac{6K}{a^2} + \frac{6}{N^2}\left(\frac{\da^2}{a^2} + \frac{\ddot{a}}{a} - \frac{\da}{a}\frac{\dot{N}}{N}\right)\,.
\end{equation}
Due to diffeomorphism invariance, the lapse function $N(t)$ can be eliminated by setting $N=1$, i.e. if we simply replace the (coordinate) time $\dot{( )}=\frac{d}{dt}$ by $\dot{( )}=\frac{d}{d \cT}$, where $d \cT=N(t)dt$ denotes a unique cosmic (proper) time. Remarkably, such redefinition does not change three-dimensional foliation of spacetime by slices of constant time. However, some other choices can be useful. For example, $N(\tau) =a(\tau)$ provides a conformal time while $dn=Hd\cT=d\ln a$ is related with an e-fold number. We stress that in MSS formulation, the lapse function turns out to play a very important role.

 Usually, one just assumes the form of stress-energy tensor for perfect fluid represented by the energy density $\rho$ and the pressure $p$ to be\footnote{In fact, one assumes a sum of noninteracting species $\rho =\sum_i \rho_i, p=\sum_i p_i$ with different barotropic factors $p_i=w_i\rho_i$.}:
\begin{equation}\label{emtensor}
T_{\mu\nu} = (p + \rho)u_\mu u_\nu + p g_{\mu\nu},
\end{equation}
with a comoving observer velocity $u^\mu = \left(\frac{1}{N},0,0,0\right)$. It remains unclear from what Lagrangian it might be obtained. We know, however, that in the most general case, the stress-energy tensor is not conserved, unless there is no anomalous coupling between the matter part of the action and the scalar field:
\begin{equation}\label{conservation}
\nabla_\mu T^{\mu\nu} = \alpha'(\Phi) T \partial^\nu \Phi.
\end{equation}
Then substituting FLRW metric
  (assuming that $\Phi = \Phi(t)$ and a barotropic perfect fluid $p = w\rho$ as a source) one gets the corresponding set of equations for the dynamical variables $a,\Phi$ (cf. \cite{borowiec2020}):
\begin{subequations}
	\begin{align}
	    \begin{split}
	        & 3H^2  = \frac{\kappa^2 N^2\,\rho}{\mathcal{A}(\Phi)} - \frac{3N^2\,K}{a^2} + \frac{1}{2}\frac{\mathcal{B}(\Phi)}{\mathcal{A}(\Phi)}\dot{\Phi}^2 - 3\frac{\mathcal{A}'(\Phi)}{\mathcal{A}(\Phi)}H\dot{\Phi}\\
	        &\quad\quad+ \frac{N^2\,}{2}\frac{\mathcal{V}(\Phi)}{\mathcal{A}(\Phi)}\,, \label{ee1}
	    \end{split}\\
	    \begin{split}
	        & 2\dot{H} + 3H^2 -2\frac{\dot{N}}{N} H = -  \frac{w N^2\,\kappa^2\rho}{\mathcal{A}(\Phi)} - \frac{N^2\,K}{a^2}\\
	        &\quad\quad-\frac{\mathcal{B}(\Phi) + 2\mathcal{A}''(\Phi)}{2\,\mathcal{A}(\Phi)} \dot{\Phi}^2  + \frac{N^2\,\mathcal{V}(\Phi)}{2\,\mathcal{A}(\Phi)}\\
	        &\quad\quad- \frac{\mathcal{A}'(\Phi)}{\mathcal{A}(\Phi)}\left(2H\dot\Phi + \ddot{\Phi}-\frac{\dot{N}}{N}\dot{\Phi}\right)\, ,\label{ee2}
	    \end{split}\\
		\begin{split}
			& \left(3(\mathcal{A}'(\Phi))^2 + 2\mathcal{A}(\Phi)\mathcal{B}(\Phi)\right)(\ddot{\Phi} -\frac{\dot{N}}{N}\dot{\Phi}) = -3(3(\mathcal{A}'(\Phi))^2 \\
			&+ 2\mathcal{A}(\Phi)\mathcal{B}(\Phi))H\dot{\Phi}-\left((\mathcal{A}(\Phi)\mathcal{B}(\Phi))' + 3\mathcal{A}'(\Phi)\mathcal{A}''(\Phi)\right)\dot{\Phi}^2\\
			&\quad\quad+ N^2\left(2\mathcal{V}(\Phi)\mathcal{A}'(\Phi) - \mathcal{V}'(\Phi)\mathcal{A}(\Phi)\right) \\
			& \quad\quad + N^2\,\kappa^2\rho(1 - 3w)\left[\mathcal{A}'(\Phi) - 2\alpha'(\Phi)\mathcal{A}(\Phi)\right]\,. \label{ee3}
		\end{split}
	\end{align}
\end{subequations}
Through this paper,  prime $( )^\prime=\frac{d}{d\Phi}$ denotes differentiation w.r.t. the current scalar field, overdot  $\dot{( )}=\frac{d}{dt}$ - w.r.t. the current coordinate time corresponding to generic lapse function $N(t)dt=d\cT$, and $H = \dot{a}/a$ the Hubble parameter as measured by the coordinate observer. An observable (as measured by comoving observer) Hubble parameter $\cH=\frac{H}{N}$ is commonly used in cosmology. In fact, equations \eqref{ee1}-\eqref{ee3} can be expressed in terms of gauge invariant quantities as $N^{-1}H, N^{-1}\dot\Phi, N^{-2}(\ddot{\Phi} -\frac{\dot{N}}{N}\dot{\Phi}),\ldots$, i.e. they are equivalent to the equations obtained by setting $N=1, t=\cT$.

For the perfect fluid stress-energy tensor (\ref{emtensor}), after setting $\nu = 0$ in (\ref{conservation}), we get:
\begin{equation}
\dot{\rho} + 3H(p + \rho) = -\dot{\alpha}(3p - \rho),
\end{equation}
which is valid for arbitrary local time variable $t$. 
It admits the following solution:
\begin{equation}\label{Tdensity}
\rho(a,\Phi) = \sum_i \rho_{0,i}a^{-3(1+w_i)}e^{(1-3w_i)\alpha(\Phi)},
\end{equation}
where $i$ labels different components of energy density content. We also assumed that, for each component, the relation between energy density and pressure is given 
by: $p_i = w_i\rho_i$. The index '0' denotes unknown value of the density component $\rho_{0,i}$ at the epoch when $a=1$\footnote{In standard cosmology ($\alpha=0$),  after normalization of the scale factor $a\in (0, 1]$, this is a present-day epoch.  Thus, in cosmological applications, initial data are replaced by present-day data.}.  This solution depends on time implicitly only throughout the dynamical variables $(a,\Phi)$. Due to the coupling $\alpha(\Phi)$, it is possible to relate the effective mass of the scalar field to the density of matter, so that its range is influenced by local environment. Such a process, called the 'Chameleon mechanism', is responsible for hiding the effects of a massive scalar field on scales smaller than the Solar System, where it should be possible to detect its presence \cite{khoury}.

\subsection{MSS reformulation}

In order to analyze different theories using methods of dynamical systems, we need to obtain from the action (\ref{action1}), the effective Lagrangian employing only the time-dependent variables. This is known as a minisuperspace formalism (there is a wide literature on the subject, see e.g. \cite{mss06}-\cite{Barrow2018} and references therein)  which, upon substitution of the metric (\ref{metric:metric1}) and integrating over the spatial variables reads as: 
\begin{equation}\label{effaction}
	\begin{split}
		S_\text{MSS}[a,\Phi] =&\frac{1}{2\kappa^2}\int dt   \Big[\frac{1}{N}\left(-6a\ca \da^2 - 6a^2 \ca' \da \dot{\Phi} + a^3\cb \dot{\Phi}^2\right) \\
		&- N V_{\text{MSS}}\Big],
	\end{split}
\end{equation}  
where:
\begin{equation}\label{Vmss}
	V_\text{MSS}(a, \Phi) = -6K a \ca(\Phi) + a^3\cv(\Phi) + V_\text{matter}(a, \Phi, \chi)\,,
\end{equation}
where $V_\text{matter}(a, \Phi, \chi) = - 2\kappa^2 L_\text{matter}(a, \Phi, \chi)$.
The effective MSS Lagrangian: 
\begin{equation}\label{effl}
	L_\text{MSS}(N, x, \dot{x}) = \frac{1}{2N}m_{jk}(x)\dot{x}^j\dot{x}^k - N V_\text{MSS}(x)
\end{equation}
lives in a three-dimensional configuration space, where $(x^j)|_{j = 1,2} = (a, \Phi)$, the dot denotes differentiation w.r.t. the (local) Newtonian time $t$. 
In fact, due to non-dynamical character of the variable $N$, for which the momentum $p_N\doteq \frac{\partial L_\text{MSS}}{\partial \dot{N}}=0$, the Lagrangian \eqref{effl} is singular and can be reduced to the plane as a configuration space.
The kinetic energy term of such reduced system is determined by 
a metric  $m_{ij}(x),\: i,j = 1,2$:
\begin{equation}\label{metric2d}
	m_{ij}
	\equiv m_{ij}(a,\Phi) =  
	\begin{pmatrix}
		-12a\ca(\Phi) & -6a^2\ca'(\Phi) \\
		-6a^2\ca'(\Phi) & 2a^3\cb(\Phi) \\
	\end{pmatrix}\,.
\end{equation}
providing the geometry to a two-dimensional configuration $(a,\Phi)$-plane $\subset\mathbb{R}_+\times \mathbb{R}$ a.k.a. MSS. 
This metric is proportional to the Hessian matrix $\frac{\partial^2 L_\text{MSS}}{\partial \dot{x}^i\partial \dot{x}^j}$.
Thus the action (\ref{effaction}) describes a motion of single particle in  two-dimensional configuration space; the space itself is Lorentzian or Euclidean  manifold equipped with the metric \eqref{metric2d}. We remark that the coordinate $\Phi$ should belong to a maximal common domain of all functions $\{\ca, \cb, \cv, \alpha\}$ determining  the frame.

The first difficulty we encounter is writing an effective Lagrangian for the matter source $T_{\mu\nu}$. However, using the solution \eqref{Tdensity}, this task simplifies to provide \footnote{Let us remember that $L_\text{matter}$ is the Lagrangian density, which means it already contains the determinant of the spacetime metric \eqref{metric:metric1}.}
 \begin{equation}\label{ml}
 \begin{split}
     V_\text{matter}(a,\Phi)=& -2\kappa^2 L_\text{matter}(a,\Phi) = 2\kappa^2\,\sqrt{-g}T_{00} \\
     &=2\kappa^2\, a^3 \rho \equiv2\kappa^2\, \sum_i \rho_{i,0}a^{-3w_i}e^{(1-3w_i)\alpha(\Phi)}
 \end{split}
 \end{equation}
 the term contributing to the total MSS potential \eqref{Vmss}. Curiously, the presence of non-minimally coupled `cosmic strings` ($w=-1/3$, $\alpha(\Phi)=1/2\ln \ca(\Phi)$) in \eqref{ml} can cancel out the effect of spacial curvature term $-6K a \ca(\Phi)$ for $K>0$ in \eqref{Vmss} and solves so-called flatness problem.
 
 We are ready now to calculate Euler-Lagrange equations of motion for \eqref{effl}. Since the lapse function  $N$ enters the action in a non-dynamical way, providing a constraint equation, which can be obtained from Euler-Lagrange equations:
\begin{equation}
\frac{\delta L_\text{MSS}}{\delta N}\equiv \frac{\partial L_\text{MSS}}{\partial N} - \frac{d}{dt}\frac{\partial L_\text{MSS}}{\partial \dot{N}} \,=\,\frac{\partial L_\text{MSS}}{\partial N} = 0,
\end{equation} 
that is equivalent to the condition:
\begin{equation}\label{c1}
\frac{1}{2N^2}m_{ij}\dot{x}^i\dot{x}^j + V_\text{MSS} = 0\,.
\end{equation}
After choosing the gauge $N = \text{const}=1$, the last equation  gives, on one hand, the Friedmann equation for a theory with a scalar field (cf. \eqref{ee1}):
\begin{equation}\label{friedmann}
3\ch^2 = -3\frac{\ca'}{\ca}\ch\dot{\Phi} + \frac{\cb}{2\ca}\dot{\Phi}^2-\frac{3K}{a^2} + \frac{\cv}{2\ca} + \frac{V_\text{matter}}{2 a^3 \ca}\,.
\end{equation}
On the other hand it represents zero Hamiltonian energy condition for the reduced Lagrangian system  on a plane.
The Euler-Lagrange equations of motion for the remaining two variables can be now easily obtained: 
\begin{equation}\label{eom1}
\begin{split}
    m_{il}\ddot{x}^l +& \frac{1}{2}\delta_i^p(\partial_j m_{pk}+\partial_k m_{pj}-\partial_p m_{jk})\dot{x}^j\dot{x}^k\\
    &=m_{il}\frac{\dot{N}}{N}\dot{x}^l -N^2 \partial_i V_\text{MSS}.
\end{split}
\end{equation}
In such approach the lapse function $N(t)$ plays a role of an additional gauge degree of freedom which is responsible for a time re-parameterization and suitably modifies the constraint \eqref{c1} \footnote{We stress that $\dot{() }=\frac{d}{dt}()$ denotes the differentiation with respect to the local (coordinate observer) time $dt=d\cT/N$.
The case N=1 corresponds to the cosmic time $\cT$ and the Friedmann equation  \eqref{ee1}.} 
\footnote{For $N=const$, a mechanical system represented by the Lagrangian \eqref{effl} is conservative, i.e. the Hamiltonian energy $\fH=\frac{1}{2N}m_{jk}\dot{x}^j\dot{x}^k + N V_\text{MSS}(x)$ is conserved, cf. \eqref{c1}.}.

It is worth underlying that the above set of equations (including \eqref{c1}) which are obtained for a constrained Newtonian mechanical system represented by the effective MSS Lagrangian \eqref{effl} is fully equivalent to the ones obtained from the Einstein field equations imposed on the FLRW metric \eqref{metric:metric1}. Without taking into account the lapse function it would not be possible since equations \eqref{eom1} themselves are not the same as \eqref{ee1}-\eqref{ee3}. 

 A more familiar form of \eqref{eom1} can be reached by an assumption that the metric \eqref{metric2d} is reversible, i.e.  its determinant:
 \begin{equation}\label{det1}
 m:=\det (m_{ij})=-12a^4 \mathcal{I}\,,\qquad \mathcal{I}=2\ca \cb +3(\ca')^2	
 \end{equation}
 should not identically vanish, therefore the case $2\ca \cb +3(\ca')^2\equiv 0$ has to be excluded. \footnote{This excludes, e.g., Palatini $f(R)$ gravity, for which the Lagrangian \eqref{effl} for $N=1$ is still singular.}
 \footnote{Singular points $\Phi_s$ such that  $(2\ca \cb +3(\ca')^2)|_{\Phi=\Phi_s}=0$, which generate singular lines in MSS, could be allowed. There is also a Big Bang type singularity at the origin $a=0$, cf. \eqref{det1}. Notice that around singular points the equation \eqref{eom2} should be replaced by more general \eqref{eom1}.}
 In such a case  one gets:
 \begin{equation}\label{eom2}
\ddot{x}^i + G^i_{jk}\dot{x}^j\dot{x}^k =\frac{\dot{N}}{N}\dot{x}^i -N^2 m^{ij}\partial_j V_\text{MSS},
\end{equation}
where $m^{ij}$ denotes the inverse metric and $G^i_{\:kl}$ are Christoffel symbols:  
\begin{equation}
\begin{split}
    	 & m^{ij} = -\frac{1}{12 a^4 \ci}
	\begin{pmatrix}2a^3\cb
		& 6a^2\ca' \\
		6a^2\ca' & -12a\ca \\
	\end{pmatrix}, \\
	& G^i_{\:kl} = \frac{1}{2}m^{ip}\left(\partial_k m_{lp} + \partial_l m_{kp} - \partial_p m_{kl}\right)\,.
\end{split}
\end{equation}
To be precise, one should distinguish two cases: $\mathcal{I}> 0$ when the MSS metric has Lorentzian signature and the opposite one with Euclidean signature. The latter one results from 'incorrect' kinetic energy sign in the Einstein frame (see below) and is related to the presence of a so-called ghost scalar field \cite{ghost}. The change of signature through a singular point $\mathcal{I}= 0$, if possible, could also be of some interest. 

Let us notice that the Eq.(\ref{eom2}) determines a geodesic trajectory if  $V_\text{MSS}=\text{const}$ \footnote{$\dot{N}=0$ means that the cosmic time parametrizes the temporal component and it is the geodesic parameter.}. 
It occurs when there is no spatial curvature $K$, self-interaction potential of the scalar field is constant (playing the role of cosmological constant) and the only allowed form of matter is dust, since for this particular type of matter, if there is no anomalous coupling between the scalar field and matter (or the coupling amounts to multiplying the energy density by a number), $L_\text{matter} = \text{const}$.
In general, the r.h.s. represents, besides the time reparametrization term, a Newtonian potential force obtained from: $V_\text{MSS}(a,\Phi)$. 
 More explicitly, the equations of motion for $a$ and $\Phi$ obtained from (\ref{eom2}) take the following form 
\begin{subequations}
	\begin{align}
		\begin{split}
			\frac{\ddot{a}}{a}  = &- \frac{\ca\cb + 3(\ca')^2}{\mathcal{I}}\left(\left(\frac{\da}{a}\right)^2+ \frac{N^2 K}{a^2}\right)+\frac{\ca'\cb}{\mathcal{I}}\frac{\da}{a}\dot{\Phi}\\
			&\quad-\frac{\cb^2-\ca'\cb'+2\ca''\cb}{2\mathcal{I}}\dph^2+\,\frac{\dot{N}}{N}\frac{\da}{a} \\
			&\quad  + \frac{N^2}{2\ci}\Bigg[\cb\cv + \ca'\cv' + \left[\ca'(1-3w)\alpha'-w\cb\,\right]\rho_w\Bigg],
		\end{split}  \label{ev1}\\
		\begin{split}
			\ddot{\Phi} = & \frac{3\ca\ca'}{\mathcal{I}}\left(\left(\frac{\da}{a}\right)^2 +\frac{N^2 K}{a^2}\right)- \frac{6\ca\cb+6(\ca')^2}{\mathcal{I}}\frac{\da}{a}\dph\\
			&\quad-\frac{3\ca'\cb+2\ca\cb'+6\ca'\ca''}{2\mathcal{I}}\dph^2+ \frac{\dot{N}}{N}\dph \\
			& \quad + \frac{N^2}{2\ci} 
			\Bigg[3\ca'\cv -2\ca\cv' -[3w \ca' + 2\ca (1-3w)\alpha'\,]\rho_w\Bigg]\,,\label{ev2}
		\end{split}
	\end{align}
\end{subequations}
where $\rho_w=2\kappa^2\rho_{0,w} a^{-3(w+1)}\,\exp{(1-3w)\alpha(\Phi)}$ represents  dimensionless density of a single  perfect fluid component with the barotropic factor $w$ or a sum over all barotropic components otherwise.\footnote{Further, in order to simplify the notation, we set $2\kappa^2=1$ or incorporate it into  $\rho_{0,w}\mapsto 2\kappa^2\rho_{0,w}$.} 
As already mentioned,  we are interested in solutions satisfying the zero (Hamiltonian) energy  constraint \eqref{ee1}, where $H=\frac{\dot a}{a}$.
The above formulas significantly are simplified in the case of dust matter $w=0$, and/or minimal coupling $\alpha'=0$.

\subsection{Cauchy data}

In order to solve the above, constrained system of second-order ODE one needs to choose the gauge and some initial data.  Instead, in cosmology, we are forced to use present-day astrophysical data extracted from observation. Let us then discuss how is possible to replace unknown present-day values $(a_0, \Phi_0, \dot a_0, \dot\Phi_0)$, at the cosmic time $\cT_0$ being the age of our universe, by viable observational data. First of all, for the cosmic time, $N=1, H=\cH$ and the Hamiltonian energy \eqref{friedmann} is conserved along each trajectory. This implies that zero energy level is preserved as well and has to be used to constrain present-day data.
(In fact, the only way a condition \eqref{friedmann} infers the system \eqref{ev1}-\eqref{ev2} is through imposing constraints on the initial data.) 
For further applications it is convenient to replace unknown constants $\rho_{i,0}$  in \eqref{ml} by 
dimensionless densities $\Omega_{i,0}$:
\begin{equation}\label{ml1}
	V_\text{matter}(a,\Phi) =3\cH_0^2\sum_i\Omega_{0,i}a^{-3w_i}\,e^{(1-3w_i)\alpha(\Phi)}\,
\end{equation}
that are constrained by observations \cite{planck2015}, where:
\begin{eqnarray}
	\Omega_{i,0}=\frac{2\kappa^2\rho_{0,i}}{3\cH_0^2}\,.
\end{eqnarray}
Here we used a usual textbook definition of $\Omega_{i,0}$. 

As customary, we firstly normalize the scale factor $a$ in such a way that $a_0=1$. It is possible since the derivatives of $a(t)$ are present in scale-invariant form $\dot{a}/a, \ddot{a}/a$ in eq.s \eqref{ev1}-\eqref{ev2}, as well as in the constraint condition \eqref{friedmann}.
Then $\dot a_0=\cH_0$ takes the present day value of Hubble parameter. 
Therefore the values $(\Phi_0, \dot\Phi_0)$ are constrained by the Hubble law \eqref{friedmann}
\begin{equation}\label{friedmann2}
\begin{split}
    	3\cH_0^2 = &-3\frac{\ca'}{\ca}\Big|_{\Phi=\Phi_0}\, \cH_0\dot{\Phi}_0 + \frac{\cb}{2\ca}\Big|_{\Phi=\Phi_0} \,\dot{\Phi}_0^2 - K + \frac{\cv}{2\ca}\Big|_{\Phi=\Phi_0}\\
    	&+ \frac{V_\text{matter}}{2\,\ca}\Big|_{a=1, \Phi=\Phi_0}
\end{split}
\end{equation}
 being a quadratic equation for $\dot{\Phi}_0$. Of course, the scalar field self-interacting potential $\cv(\Phi)$ should possess good inflationary properties, cf. \cite{martin2014}.\\  
{\bf Remark:} 
Assuming further that the scalar field has no dynamics at the present epoch, i.e. $\dot{\Phi}_0=0$, 
one gets $\Lambda$CDM type relation
\begin{equation}\label{friedmann3}
	1 =   \Omega_\Lambda + \Omega_K +\frac{1}{2\ca (\Phi_0) } \sum_i\Omega_{0,i} \,e^{(1-3w_i)\alpha(\Phi_0)}\,\,,
\end{equation}
where $\Omega_K=-\frac{K}{3\cH^2_0}$ and $\Omega_\Lambda=\frac{\cv (\Phi_0)}{6\cH_0^2\ca (\Phi_0)}$ could play a role of cosmological constant. 
In such scenario the observed matter $\widetilde \Omega_{0,i}$ could differ from 'true' matter $ \Omega_{0,i}$ by a factor $\frac{\exp{((1-3w_i)\alpha(\Phi_0))}}{2\ca (\Phi_0) }$. In fact, this factor does not depend on a numerical value of $\cH_0$
\footnote{Therefore, recently reported tensions between different $\cH_0$ measurements, see e.g.  \cite{h01,h02} (cf. also \cite{not_h01,not_h02} for an alternative explanations), are irrelevant in these considerations.}. Some examples will be considered later on.

\subsection{More on  
Lagrangian systems with zero energy constraints}

The absence of $\frac{\dot{N}}{N}\dot{x}^i$ term in \eqref{eom2} means that a local time $t$ is proportional to  the cosmological time $\cT$. However, the equations \eqref{eom2} are still valid if we assume implicit time-dependence of the lapse function, 
 i.e.:
\begin{equation}\label{laps2}
	N(t)=N(a(t),\Phi(t))\,,\qquad \dot N=\dot a\partial_a N +\dot\Phi\partial_\Phi N\,.
\end{equation} 
 In such case, the equation \eqref{eom2} can be recast into a new equivalent form
  \begin{equation}\label{eom2b}
 	\ddot{x}^i + \widehat{^N\! G}^i_{jk}\dot{x}^j\dot{x}^k =  -N^2 m^{ij}\partial_j V_\text{eff},
 \end{equation}
where $\widehat{^N\!G}^i_{\:kl}=G^i_{\:kl}-\delta^i_{(k}\partial_{l)}\ln N$ are  components of a projectively equivalent  connection which in general is not  metric \footnote{We recall that two metrics are called projectively equivalent if they have the same geodesics as un-parametrized curves. The metricity of such connection is related with Lie problem and bihamiltonian St\"ackel systems, see e.g. \cite{Ibort,Manno}.}. Now the zero Hamiltonian energy constraint \eqref{c1} has to be imposed by hand.
 
Alternatively, if one incorporates the lapse function into the metric 
$m_{ij}\mapsto \widetilde{^N\!m}_{ij}\equiv\frac{1}{N(a, \Phi)\,}m_{ij}(a, \Phi)$ and into the potential $V_\text{MSS}\mapsto \widetilde{^N\!V}_\text{MSS}\equiv N(a,\Phi)\,V_\text{MSS}(a, \Phi)$, then the constrained system \eqref{c1},\eqref{eom2} can be replaced by the equivalent one 
\begin{equation}\label{eom2c}
\begin{split}
    & \ddot{x}^i + \widetilde{^N\!G}^i_{jk}\dot{x}^j\dot{x}^k =  -\,\, \widetilde{^N\! m}^{ij}\partial_j \widetilde{^N\!V}_\text{MSS},\\
    & \frac{1}{2}\,\widetilde{^N\! m}_{ij}\dot{x}^i\dot{x}^j + \widetilde{^N\!V}_\text{MSS} = 0\,,
\end{split}
\end{equation}
where  $\widetilde{^NG}^i_{\:kl}=G^i_{\:kl}-\delta^i_{(k}\partial_{l)}\ln N+{1\over 2}m_{kl}m^{ij}\partial_j\ln N$ are  components of the Levi-Civita connection of 
the conformally related metric $\widetilde{^N\!m}_{ij}$.
Now, in contrast to the previous case, the first equation comes from the non-singular
Lagrangian system $\widetilde{^N\!L}_\text{MSS}=\frac{1}{2}\,\widetilde{^N\! m}_{ij}\dot{x}^i\dot{x}^j - \widetilde{^N\!V}_\text{MSS}$. This shows that any choice of a gauge \eqref{laps2} provides an alternative classical description of the original constrained system.
In particular, for $N=1/V_\text{MSS}$ one gets Lagrangian with purely kinetic energy (cf. \cite{Jacobi}).
 
 Before discussing further some particular choices for the lapse function \eqref{laps2} we want to extend the techniques known for kinetic energy  Lagrangians, see e.g. \cite{Ibort,Manno} and references therein, to our case including potential energy. In this scenario, one makes the following choice $d\cT=N(y)dy$, where the function  
$N(y)$ has to be determined and $y=\left(x^i\right)|_{i=2}$ is the second coordinate, while $x = \left(x^i\right)|_{i=1}$ is the first coordinate. Now (cf. \eqref{eom2b}, $\dot{()}\equiv \frac{d}{d y}$ and $\dot{y}=1$):
\begin{eqnarray*}
	\ddot{x}\: +\: \widehat G^1_{11}\dot{x}^2 \:+\: \widehat G^1_{12}\dot{x}+ \widehat G^1_{22}=  -N^2 m^{1j}\partial_j V_\text{MSS},    \\
	\widehat G^2_{11}\dot{x}^2\: +\widehat G^2_{12}\dot{x}+\widehat G^2_{22} = -N^2 m^{2j}\partial_j V_\text{MSS},   
\end{eqnarray*}
where $\widehat{G}^i_{\:kl}=\widehat{^{N} G}^i_{\:kl} $ for simplicity.
These two can  be rearranged into a single equation:
\begin{equation}
\begin{split}
    \ddot{x} & \: = \: \widehat G^2_{11}\dot{x}^3 \:-\: (\widehat G^1_{11}-2\widehat G^2_{21})\dot{x}^2- \widehat G^1_{22} -N^2 m^{1j}\partial_j V_\text{MSS}\\
    & + (\widehat G^2_{22}-2\widehat G^1_{21} +N^2 m^{2j}\partial_j V_\text{MSS})\dot{x}\,.
\end{split}
\end{equation}
In order to make this equation self-consistent, one needs to specify $N(y)$. To this aim we use the constraint equation \eqref{c1}
from which one gets:
\begin{equation}
	N^2  =-\frac{m_{ij}\dot{x}^i\dot{x}^j }{2V_\text{MSS}}=-\frac{m_{11}\dot{x}^2+2 m_{12}\dot{x}+m_{22}}{2V_\text{MSS}}
\end{equation}
Substituting back to the last equation yields
\footnote{Notice, that elements $\widehat G^2_{11}=G^2_{11}, \widehat G^1_{11}-2\widehat G^2_{21}= G^1_{11}-2 G^2_{21}, \widehat G^2_{22}-2\widehat G^1_{21}= G^2_{22}-2 G^1_{21},\widehat G^1_{22}=G^1_{22}$ 
are invariant with respect to the projective transformation $G^k_{ij}=\widehat G^k_{ij}+\delta^k_i \omega_j+\delta^k_j \omega_i$, preserving unparametrized geodesics for any covector $\omega_i$.}:
\begin{equation}
\begin{split}
    \ddot{x}\: = & \: (G^2_{11}+\frac{1}{2}m_{11} m^{2j}\partial_j \ln V_\text{MSS})\dot{x}^3\\
    &\:+\: \left(2G^2_{21}-G^1_{11}+(\frac{1}{2}m_{11} m^{1j} -m_{12} m^{2j})\partial_j \ln V_\text{MSS}\right)\dot{x}^2\\ & +\left(G^2_{22}-2G^1_{21}
	+(\frac{1}{2}m_{22} m^{2j}-m_{12} m^{1j})\partial_j \ln V_\text{MSS} \right)\dot{x}\\
	& - G^1_{22} +\frac{1}{2}m_{22} m^{1j}\partial_j \ln V_\text{MSS}\,.
\end{split}
\end{equation}
Interchanging indices $1\leftrightarrow 2$ one can get similar expression for $y$.

As a first example, one can use so-called e-fold number $n =\ln a$ as dimensionless evolution parameter. In this case $N=H^{-1}, dn=H\,d\,T$ and $\frac{d a}{d n}=a$ one finds
\begin{equation}\label{efold}
\begin{split}
    \ddot{\Phi}\:& = \: a^{-1}\left(G^1_{22}-\frac{1}{2}m_{22} m^{1j}\partial_j \ln V_\text{MSS}\right)\dot{\Phi}^3 +\\ 
    &\left(2G^1_{12}-G^2_{22}-(m_{12} m^{1j} -\frac{1}{2}m_{22} m^{2j})\partial_j\ln V_\text{MSS}\right)\dot{\Phi}^2\,+
    \\
	& \left(G^1_{11}-2G^2_{12}+(m_{12} m^{2j} - 
	\frac{1}{2}m_{11} m^{1j})\partial_j \ln V_\text{MSS}+a^{-1}\right)a\,\dot{\Phi}\\
	& +\left( -G^2_{11} +\frac{1}{2}m_{11} m^{2j}\partial_j \ln V_\text{MSS}\right)\,a^2\,.
\end{split}
\end{equation}


\subsection{MSS Hamitonian formalism and Wheeler-DeWitt quantization}

It is well known that Hamiltonian formalism replaces second order Euler-Lagrange equations  for  the Lagrangian  $L(x, \dot{x}) = \frac{1}{2}m_{ij}\dot{x}^i\dot{x}^j -  V(x)$ on a configuration manifold by first order (autonomous) dynamical system on the corresponding phase space (a cotangent bundle) with a conserved Hamiltonian energy
$\fH(x, p) = \frac{1}{2}m^{ij}p_i p_j +  V(x)$.
This correspondence plays a fundamental role in a canonical quantization of such Lagrangian systems.  
For  constrained systems  the traditional relations between Lagrangian and Hamiltonian formalism are broken. The best way for Hamiltonian description of constrained systems is provided by the well-known Dirac formalism \cite{Dirac}. In the previous sections we discussed how the singular Lagrangian \eqref{effl} in three-dimensional configuration space $(N, a, \Phi)$ can be reduced to two-dimensional constrained system with an additional gauge freedom. 
Here, for the sake of completeness, we briefly present associated Hamiltonian description and its application to the Wheeler-DeWitt quantization procedure.
In order to warm up, one can directly check   that if one takes the Hamiltonian
\begin{equation}\label{efH}
	\fH (x^i,\ p_i, N) = \frac{N(t)}{2} m^{jk}(x)\,p_j p_k + N(t) V_\text{MSS}(x)\,,
\end{equation}
where $p_i = \frac{\partial L_\text{eff}}{\partial \dot{x}^i} = \frac{1}{N} m_{ij}\dot{x}^j$ then the Lagrangian constrained system  \eqref{c1}, \eqref{eom2} is equivalent to the first order constrained Hamiltonian one in the corresponding (reduced) phase space $(x^i, p_k)$:
\begin{equation}\label{efH1}
	\begin{cases}
		\dot{x}^i  = \frac{\partial \fH}{\partial p_i}\equiv N m^{ik}p_k\,, \\
		\dot{p}_i  = - \frac{\partial \fH}{\partial x^i}\equiv
		-\frac{N}{2}\partial_i m^{jk}p_j p_k - N \partial_i V_\text{MSS}\,, \\ 
		\frac{\partial \fH}{\partial N} = \frac{1}{N}\, \fH=0
		\Leftrightarrow \frac{1}{2}m^{jk}p_j p_k +  V_\text{MSS}=0\,.
	\end{cases}
\end{equation}
Now, changing the time variable ${d\over dt}\mapsto {d\over d\tilde{t}}={d\over Ndt}$ we are left with the Hamiltonian autonomous dynamics corresponding to $N=1$ in \eqref{efH1}. The last equation (zero energy condition) is equivalent to \eqref{c1} and, in fact, constrains the initial conditions only.

Similarly, assuming special form \eqref{laps2} of lapse function, one can always find equivalent constrained autonomous dynamical system with the Hamiltonian  (cf. \eqref{eom2c})
\begin{equation}\label{effH2}
	 \widetilde{^N\!\fH}=\frac{1}{2}\,\widetilde{^N\! m}^{ij}\widetilde{^N\!p}_i\widetilde{^N\!p}_j + \widetilde{^N\!V}_\text{MSS}\,, \qquad \widetilde{^N\!\fH}= 0\,
\end{equation}
satisfying the zero energy condition.
 
An interesting question appears now in the context of Wheeler-DeWitt MSS formalism: to what extend quantization of such classically equivalent systems provides  equivalent quantum mechanical description.  In more technical terms one can ask how much a choice of gauge $N(a,\Phi)$ changes physical output of the corresponding quantum formalism. 
In order to answer this question let us recall that the Wheeler-DeWitt canonical quantization of the original system \eqref{efH} ($N=1$) yields a Schr\"{o}dinger type (stationary) wave equation in the form (see e.g. \cite{Capo2015,Hawking,wiltshire,barvinsky})
\begin{equation}\label{WdW}
\underline{\fH}\Psi\equiv (\triangle_m + V_\text{MSS}(x)) \Psi(x)=0,
\end{equation}
with the aim to find  a wave function of the Universe $\Psi(x)$.  Here $\underline{\fH}$ stands for quantum Hamiltonian and $\triangle_m\Psi={1\over\sqrt{m}}\partial_i(\sqrt{m}\,m^{ij}\partial_j\,\Psi)$  denotes MSS Laplace-Beltrami operator 
\footnote{One should remember that canonical MSS quantization causes the problem of operator ordering in Laplace-Beltrami part  which according to Hawking and Page \cite{Hawking} can be resolved.}. 
Specifically, in two-dimensions, one has $\triangle_{\omega m}={1\over \omega}\triangle_{m}$, where $\omega(x)$ is an arbitrary conformal factor. This implies that after Wheeler-DeWitt quantization of \eqref{effH2} in different gauges
\begin{equation}\label{WdW2}
	\underline{\widetilde{^N\!\fH}}=N\,\underline{\fH}\,,
\end{equation}
the wave function $\Psi$ remains the same. This answers the questions. 
Moreover, any two-dimensional metric is conformally flat, i.e. in a suitable (isothermal, see the next section) coordinates takes a diagonal  form $ds^2= \omega(x,y)(dx^2 + \text{sign}(m)dy^2)$. Now, combining the use of these coordinates and gauge freedom $N=\omega(x,y)$ allows one to perform Wheeler-DeWitt quantization  by making use  of flat Laplacian $\partial_x^2+\text{sign}(m)\partial_y^2$ and modified potential $ V_\text{MSS}(x)\mapsto \widetilde{^\omega\!V}_\text{MSS}(x)= \omega(x)\,V_\text{MSS}(x)$.
In such way 
one can obtain a preferred operator ordering and elude the problem mentioned in \cite{Hawking}.

 \subsection{Universal coordinates}
So far we have used a natural MSS coordinate system $(a, \Phi)$ inherited from dynamical variables associated with the initial  action  \eqref{action1}. Thus the MSS Lagrangian \eqref{effl} is determined by the frame $\{\ca,\cb, \cv, \alpha\}$ functions and the matter content \eqref{ml}.
Since the Lagrangian formalism is (diffeomorphism) invariant with respect to any (local) coordinate change in a configuration MSS,  we might be tempted to use it in order to simplify calculations. On the other hand, such transformation takes us away from physical variables $(a, \Phi)$ and should be used with care.

The first idea is to diagonalize the metric $m_{ab}$ by means of a conformal factor, since every two-dimensional pseudo-Riemannian manifold is conformally flat. We start by eliminating the off-diagonal terms in the metric (\ref{metric2d}) by a simple change of variables: $a(\Phi, \xx) = \ca(\Phi)^{-1/2} \aa(\xx)$, where the function $\aa(\xx)$ is to be determined. This can be done locally, around  any nonsingular point $a>0, \Phi\neq\Phi_s$. The metric will assume the following form:
\begin{equation}\label{mss-1}
	ds^2_{(2)} = \ca^{-\frac{5}{2}}\aa^3\,\left(3\left(\ca'\right)^2 + 2\ca\cb\right) d{\Phi}^2 - 12\ca^{-\frac{1}{2}}\aa\,(\aa')^2 d{\xx}^2.
\end{equation}
It can be inferred from the transformation properties (see Appendix A for details) that the unknown function $\aa(\xx)$ is frame-independent. Moreover $\aa=a$ whenever $\ca=1$ (e.g. in Einstein frame). For these reasons, we shall call $\aa$ a universal scale factor. 

We redefine the scalar field, introducing a new one, $\psi$, called a 'universal scalar field' due to the fact that it is conformally-invariant under transformations of the space-time metric $g_{\mu\nu}$ and redefinition of $\Phi$. The field is defined as follows:\footnote{An expression under the square-root must be positive therefore $\pm$ corresponds to  $\text{sign}(\ci)=-\text{sign}(m)$, cf. \eqref{det1}. We stress that  $\text{sign}(m)$ is invariant with respect to MSS coordinate change.}
\begin{equation}\label{invariantfield}
	\frac{d\psi}{d\Phi} = \pm\sqrt{\pm\mathcal{I}\, \mathcal{A}^{-2} }.
\end{equation}
Requiring the metric  \eqref{mss-1} to be conformally flat one can solve $\aa$ to be: 
\begin{equation}\label{efold}
	\aa(\xx) \sim e^{\pm\frac{\sqrt{3} \xx}{6}}.
\end{equation}
The $\pm$ sign above is due to convention. Choosing $ \aa(\xx) = 
e^\frac{\sqrt{3} \xx}{6} $
one can interpret $\xx$ as a universal e-fold number for $\aa$. With this choice, the metric will be of the following conformally flat form:\footnote{In Riemannian geometry such kind of orthogonal coordinates are a.k.a. isothermal coordinates. They are preserved under the following coordinate change: $\tilde\psi=f(\psi+\xx)+g(\psi-\xx); \tilde \xx=f(\psi+\xx)-g(\psi-\xx)$, where $f,g$ are two differentiable functions of one variable.}
\begin{equation}\label{mss-2}
	ds^2_{(2)} = -\ca^{-\frac{1}{2}}(\psi)e^{\frac{\sqrt{3}\xx}{2}}( d\xx^2 + 	\mbox{sign}(m)\,\, d\psi^2 )\,.
\end{equation}
Remarkably, it depends only on one frame function $\ca(\psi)$ and describes the evolution of the universal scale factor $\aa$ as a function of the universal e-fold number $\xx=2\sqrt{3}\ln\aa$.
We shall call $(\xx,\psi)$ universal MSS coordinates.
 Its physical meaning will be explained in the next section. The corresponding Levi-Civita coefficients can be found in Appendix. 

The same change of variables has to be applied to the MSS effective potential (cf. \eqref{ev1},\eqref{ml}):
\begin{equation}
\begin{split}
    V_\text{MSS}(\xx, \psi) =&  -6K e^{\frac{\sqrt{3} \xx}{6}} \ca^{1\over2}(\psi) + e^{\frac{\sqrt{3} \xx}{2}} \ca^{-\frac{3}{2}}(\psi)\cv(\psi)\\
    &+ e^{-\frac{\sqrt{3} \xx\,w}{2}}\ca^{\frac{3 w}{2}}(\psi)\,e^{(1-3w)\alpha(\psi)} \rho_{0,w}  \,,
\end{split}
\end{equation}
where the last term can be a sum of terms with different barotropic factor $w$.
Finally, the MSS Lagrangian \eqref{effl} takes, in these coordinates, the form:
\begin{equation}\label{effl-u}
\begin{split}
    \mathfrak L_\text{MSS} =& -\frac{1}{2N}\ca^{-\frac{1}{2}}(\psi)e^{\frac{\sqrt{3}\xx}{2}}(\dot{\xx}^2 + \mbox{sign}(m)\,\, \dot{\psi}^2 ) \\
    &- N\,V_\text{MSS}(\xx, \psi)\,.
\end{split}
\end{equation}
Now, the constrained dynamical system \eqref{c1}, \eqref{eom2} with the Lagrangian \eqref{effl-u} can be replaced  by dynamically equivalent one with conformally rescaled MSS metric 
which turns out to be MSS flat ($N= \ca^{-\frac{1}{2}}(\psi)$, see Appendix):
\begin{equation}\label{effl-u2}
	\widetilde{\mathfrak L}_\text{MSS} = -\frac{1}{2} e^{\frac{\sqrt{3}\xx}{2}}(\dot{\xx}^2 + \mbox{sign} (m)\,\, \dot{\psi}^2 ) -\,\widetilde{V}_\text{MSS}(\xx, \psi)\,.
\end{equation}
where the potential is:  
\begin{equation}
\begin{split}
    	\widetilde V_\text{MSS}(\xx, \psi) & =  \ca^{-\frac{1}{2}}(\psi)V_\text{MSS}(\aa(\xx), \psi)\\
    	&= -6K \aa(\xx)+ \aa^3(\xx) \widetilde\cv(\psi) + \aa^3(\xx) \widetilde\rho_w 
\end{split}
 \,,\label{msspotential}
\end{equation}
and $\widetilde\rho_w=\rho_{0.w}\aa^{-3(w+1)}\,e^{(1-3w)\widetilde\alpha(\psi)}$, $\widetilde{\alpha}(\Phi) =\alpha(\Phi)-{1\over 2}\ln \ca(\Phi),$  $\widetilde{\mathcal{V}} (\psi)=\mathcal{V}(\psi)\mathcal{A}^{-2}(\psi)$.\\
Applying the formalism introduced previously, one may write down the equation of motion for $\xx$ as a function of the invariant scalar field $\psi$ (now $\dot{()}\equiv \frac{d}{d \psi}$):
\begin{equation}\label{psi}
\begin{split}
   2\ddot{\xx} & =  \text{sign}(m)\,\left( \frac{\ca'(\psi )}{2 \ca(\psi )} \,-\, \partial_\psi \ln V_\text{MSS} \right)\dot{\xx}^3 \\
	& + \left(\frac{\sqrt 3}{2}\,+\,\partial_\xx \ln V_\text{MSS}\right)\dot{\xx}^2	-\left(\frac{\ca'(\psi )}{2 \ca(\psi )} \,-\,\partial_\psi \ln V_\text{MSS}\right)\dot{\xx} \\
	& + \text{sign}(m)\,\left(\frac{\sqrt 3}{2} \,-\,\partial_\xx \ln V_\text{MSS}\right)\,. 	  
\end{split}
\end{equation}
We remark that for the purpose of the Wheeler-DeWitt quantization it would be more convenient to take $N=-\ca^{-\frac{1}{2}}(\psi)e^{\frac{\sqrt{3}\xx}{2}}$ in the formula \eqref{effl-u}.

 
\section{Solution-equivalent frames in MSS formalism}

Consider ST action functional \eqref{action1}. As it is well-known totality of all frames   
splits into solution-equivalent classes linked by the following transformations : 
\begin{subequations} \label{ap1}
	\begin{align}
		&\bar g_{\mu\nu}=e^{2\gamma(\Phi)}g_{\mu\nu}, \label{ct1}\\
		&\bar \Phi=f(\Phi).  \label{ct2}
	\end{align}
\end{subequations}
applied to the solution of field equations in a given frame $\{\mathcal{A},\mathcal{B},\mathcal{V},\alpha\}$.
It consists of conformal metric transformation implemented by an arbitrary function $\gamma(\Phi)$ \footnote{It is generally assumed that the first and second derivatives of $\bar{\gamma}$ exist.} accompanied by a redefinition of the scalar field
\footnote{This implies that the corresponding Levi-Civita connection undergoes the Weyl transformation $\bar\Gamma^\alpha_{\mu\nu}=\Gamma^\alpha_{\mu\nu}+2 \delta^\alpha_{(\mu}\partial_{\nu)}\gamma(\Phi)-g_{\mu\nu} g^{\alpha\beta}\partial_\beta\gamma(\Phi) $. }. 
Moreover, Jacobian of this transformation is allowed to be singular at some isolated points. 

Thus barred fields are solutions of corresponding equations of motion in a new frame
$\{\bar{\mathcal{A}},\bar{\mathcal{B}},\bar{\mathcal{V}},\bar\alpha\}$:
\begin{subequations} \label{ap2}
	\begin{align}
	\begin{split}
	    \bar{\mathcal{A}}(\bar{\Phi})=e^{2\,\check{\gamma}(\bar{\Phi})}\mathcal{A}(\check{f}(\bar{\Phi}))\,, \label{transf1}
	\end{split}\\
	\begin{split}
		  & \bar{\mathcal{B}}(\bar{\Phi})=e^{2\,\check{\gamma}(\bar{\Phi})}\Bigg(\Big(\frac{d\Phi}{d\bar{\Phi}}\Big)^2\mathcal{B}(\check{f}(\bar{\Phi}))-6\Big(\frac{d\check{\gamma}}{d\bar{\Phi}}\Big)^2\mathcal{A}(\check{f}(\bar{\Phi}))\\
		  &-6\frac{d\check{\gamma}}{d\bar{\Phi}}\frac{d\mathcal{A}}{d\Phi}\frac{d\Phi}{d\bar{\Phi}}\Bigg)\,,\label{transf2}  
	\end{split}\\
	\begin{split}
	    \bar{\mathcal{V}}(\bar{\Phi})=e^{4\,\check{\gamma}(\bar{\Phi})}\mathcal{V}(\check{f}(\bar{\Phi}))\,,\label{transf3}
	\end{split}\\
	\begin{split}
	    \bar{\alpha}(\bar{\Phi})=\alpha(\check{f}(\bar{\Phi}))+\check{\gamma}(\bar{\Phi})\,.\label{transf4}
	\end{split}
	\end{align}
\end{subequations}
where $\check f=f^{-1}$ denotes the inverse transformation and $\check\gamma(\bar\Phi)=-\gamma(\check f(\bar\Phi))$.  
In physical terms invariance of the metric tensor means that if observers of different conformal frames being related to each other by means of (\ref{ct1}) and (\ref{ct2}) agree on using one of the above metrics, then the distances measured by them will be the same. 

The following invariants of frame transformations \eqref{ap2} are well-known: 
\begin{subequations}
\begin{align}
     \widetilde{\alpha}(\Phi)&={1\over 2}\ln \frac{e^{2\,\alpha(\Phi)}}{\mathcal{A}(\Phi)}=\alpha(\Phi)-
	{1\over 2}\ln \ca(\Phi)\,,\label{invalph}\\
	\widetilde{\mathcal{V}} (\Phi)&=\frac{\mathcal{V}(\Phi)}{(\mathcal{A}(\Phi))^{2}}\,,\label{invV}\\
	\frac{d\psi(\Phi)}{d\Phi}&= \sqrt{\pm\frac{2\mathcal{A}(\Phi)\mathcal{B}(\Phi)+3(\mathcal{A}'(\Phi))^2}{\mathcal{A}^2(\Phi)}}.\label{invfield}
\end{align}
\end{subequations}
	
It is to be noticed that starting from an arbitrary frame $\{\mathcal{A},\mathcal{B},\mathcal{V},\alpha\}$ and choosing $\bar\Phi=\psi(\Phi)$ and {$\gamma(\psi)= -{1\over 2}\ln \ca(\Phi(\psi))$}
in (\ref{ct1})--(\ref{ct2}) we end up in Einstein frame with the following data
$\{\bar{\mathcal{A}}=1,\bar{\mathcal{B}}=\pm 1/2,\bar{\mathcal{V}}(\psi)=
\widetilde\cv(\Phi(\psi)) 
,\bar\alpha(\psi)=\widetilde\alpha(\Phi(\psi))\}$. For $\mathcal{B}\equiv 0$
one gets $\bar{\mathcal{B}}=1/2$.\footnote{If $\bar{\mathcal{B}}=0$ then such system is MSS degenerate, i.e. $\ci(\psi)=0$.} This is the canonical Einstein frame. Conversely, assume we know solution in the Einstein frame with some arbitrary self-interacting potential $\bar{\mathcal{V}}(\psi)$ and non-minimal coupling $\bar\alpha(\psi)$. Then by making use of
inverse transformation, one can get a solution in a frame  parametrized by arbitrary functions $\{\mathcal{A}, \mathcal{B}\}$ with suitably calculated $\Phi, \cv, \alpha$.

In application to ST cosmology one can observe that form of the metric \eqref{metric:metric1} remains invariant while its components transform accordingly: $N\mapsto \bar N= e^{\gamma(\Phi)} N$
and $a\mapsto \bar a= e^{\gamma(\Phi)} a$. Particularly, changing a frame we respectively change the notion of cosmic time: $d\bar \cT=e^{\gamma(\Phi)} d\cT$.  Furthermore, we can introduce invariant (universal) FLRW metric
\begin{equation}\label{FRW-E}
\begin{split}
    \widetilde g_{\mu\nu} &= \mathcal{A}(\Phi)\,g_{\mu\nu}\\
    &=\text{diag}\left(-1\,,  \frac{\aa^2(\tT) }{1 - K r^2}\,, \aa^2(\tT) r^2\,,\aa^2(\tT) r^2 \sin^2\theta\right),
\end{split}
\end{equation}
with invariant (universal) cosmic time $\mathfrak t$ and universal scale factor $\aa$  
(invariance of this metric simply means that 
$\bar{\mathcal{A}}\,\bar g_{\mu\nu}=\mathcal{A}\,g_{\mu\nu}$). Comoving spacial coordinates $(r,\theta,\phi)$ remain unchanged.

More explicitly, the equations of motion for $\aa$ and $\psi$ obtained from (\ref{FRW-E}), cf. \eqref{ev1}-\eqref{ev2},  take the following form 
($\text{sign}(\mathcal{I})=\pm 1$, $\dot{( ) }\equiv \frac{d}{d \tT}, \,\, N(t)dt=d\,\tT$, 
$( )'\equiv \frac{d}{d\,\psi}$):
\begin{subequations}
	\begin{align}
		\begin{split}
			\frac{\ddot{\aa}}{\aa}  = &- \frac{1}{2}\left(\left(
			\frac{\dot{\aa}}{\aa}\right)^2+ \frac{N^2 K}{\aa^2}\right) -\frac{\text{sign}(\ci)}{8}\,\dot{\psi}^2 +\,\frac{\dot{N}}{N}\frac{\dot{\aa}}{\aa}\\
			&+ \frac{N^2}{4}\Bigg[\widetilde\cv - w \widetilde\rho_w\Bigg]\,,\label{ev1u}
		\end{split}  \\
		\begin{split}
			\ddot{\psi} = & - 3\frac{\dot{\aa}}{\aa}\dot{\psi}   
			+ \frac{\dot{N}}{N}\dot{\psi} - N^2\,\text{sign}(\ci)\, \left[\widetilde\cv' +  
			(1-3w)\widetilde\alpha'\,\widetilde\rho_w\right]\,,\label{ev2u}
		\end{split}
	\end{align}
\end{subequations}
where $\widetilde\rho_w= \rho_{0.w} \aa^{-3(w+1)}e^{(1-3w)\widetilde\alpha(\Phi)}$ represents  dimensionless density of a single perfect fluid component with the barotropic factor $w$ or a sum over all barotropic components otherwise.
It turns out that all quantities in \eqref{ev1u}-\eqref{ev2u}
have invariant (frame independent) meaning, cf. \eqref{invalph}-\eqref{invfield}. Moreover, natural coordinates in Einstein frame $(\aa,\psi)$ become, after the substitution $\aa(\xx)=e^{\pm\frac{\sqrt{3} \xx}{6}}$, just universal coordinates $(\xx,\psi)$. The cosmic time in this frame has also invariant meaning. Choosing $\{\ca,\cb\}$ with the proper value of $\text{sign}(\ci)$ one can reconstruct original model.

\subsection{$f(R)$ theories of gravity}


It is well-known that metric $f(R)$ theories of gravity have a scalar-tensor representation, which can be achieved by means of a Legendre transformation. We will present here only the final result, as the process of deriving the ST representation can be found in many papers, for example, in \cite{capo2011}: 
\begin{equation}
\begin{split}
	S[g_{\mu\nu}, \Phi, \chi] = & \frac{1}{2\kappa^2}\int_\Omega d^4x \sqrt{-g}[\phi R(g) - V_{f(R)}(\Phi)] \\
	&+ S_\text{matter}[g_{\mu\nu}, \chi],
\end{split}
\end{equation}
where $V_{f(R)}(\Phi) =  \Phi R(\Phi)-f(R(\Phi)),\, \ \Phi=f'(R)$ \footnote{
There is no one-to-one correspondence between the functions $f(R)$ and $V_{f(R)}(\Phi)$. In fact $V_{f(R)}(\Phi)=\int g(\Phi)d\Phi$ depends on a choice of some  (local) inverse function $R=g(\Phi)$, where $f'(g(\Phi))=\Phi$, cf. Appendix A in \cite{kozak2019}.}.
The four functions of the scalar field we identified at the beginning of the paper, present in action (\ref{action1}), are:
$$\ca(\Phi) = \Phi, \quad\cb(\Phi) = 0, \quad\cv(\Phi) = V_{f(R)}(\Phi),\quad\alpha(\Phi) = 0.$$
The metric $m_{ij}$ on the mini-superspace is:
\begin{equation}
	m = 
	\begin{pmatrix}
		-12 a\Phi & -6a^2 \\
		-6a^2 & 0
	\end{pmatrix},
\end{equation}
which, upon diagonalization, yields the following Lagrangian density:
\begin{equation}\label{diag1}
\begin{split}
   L_\text{eff}(\psi, \xx) &=  \frac{1}{N}e^{\frac{\sqrt{3}(-\psi + 3\xx)}{6}}(- \dot{\xx}^2 + \dot{\psi}^2 ) \\
   &- N\left(-6K e^\frac{\sqrt{3}(\psi + \xx)}{6} + e^\frac{\sqrt{3}( \xx - \psi)}{2}V_{f(R)}(\psi)\right), 
\end{split}
\end{equation}
with the following redefinition of the variables:
\begin{equation}
	\begin{cases}
		\Phi = &  e^{\frac{\sqrt{3}}{3}\psi}, \\
		a = & e^{\frac{\sqrt{3}(\xx - \psi)}{6}}.
	\end{cases}
\end{equation}
For $N=1$ and the metric present in (\ref{diag1}), the Christoffel symbols are the following:
\begin{equation}\label{christfr}
\begin{split}
   & G^1_{11} = G^1_{22} = G^2_{12} = G^2_{21} = \frac{\sqrt{3}}{4}, \\
   & G^1_{12} = G^1_{21} = G^2_{11} = G^2_{22} =- \frac{\sqrt{3}}{12},
\end{split}
\end{equation}
which means that the curvature on the mini-superspace  vanishes. 

\subsubsection*{Class of $f(R)$ theories with a first-order integral of motion}
 Let us now consider a spacetime with vanishing spatial curvature. The function $N$ in the denominator of the kinetic coupling of \eqref{diag1} remains unspecified, and can be treated as a function of the MSS coordinates - $\psi$ and $x$ (which, in turn, depend on the parameter $t$, as shown in \eqref{laps2}). By performing a coordinate transformation (which is not canonical), we are able to bring the metric to the Lorentzian form. This is, of course, not possible for every theory, unless we exercise the freedom given by arbitrariness of the gauge in the theory defined on a 4-dimensional manifold, and fix the lapse function in such a way that the conformal factor multiplying the metric will be brought to a constant. Such a procedure will change the geometric properties of MSS, but the formalism defined on that space has only operational meaning, with no physical relevance. Fixing the lapse will be then equivalent to a conformal transformation of the MSS metric. 

For the metric case, let us notice that after choosing the following lapse:
  \begin{eqnarray}\label{gauge1}
      N(\psi, \xx) = 2\ca^{-\frac{1}{2}}(\Phi(\psi))\aa^3(\xx),
  \end{eqnarray}
  the potential for STT with no matter included:
\begin{equation}
\begin{split}
 V_\text{MSS}(\psi,\xx) & = a^3(\psi,\xx) \mathcal{V}(\psi) = \aa^3(\xx)\mathcal{A}^{-\frac{3}{2}}(\psi)\cv(\psi)\\
 &=e^{\frac{\sqrt{3}\xx}{2}}\mathcal{A}^{-\frac{3}{2}}(\psi)\cv(\psi),   
\end{split}
\end{equation}
is brought to the form:
\begin{equation}
U(\psi, \xx)= N(\psi, \xx) V_\text{MSS}(\psi,\xx)= 2e^{\pm\sqrt{3}\xx} \frac{\cv(\psi)}{\ca^2(\psi)}.
\end{equation}
Therefore, it becomes clear that any theory with the MSS Lagrangian of the form:
 \begin{eqnarray}\label{generalLagrangian}
     \mathcal{L}=\frac{1}{2}\left( - \dot{\xx}^2 + \dot{\psi}^2 \right) - e^{\sqrt{3}\xx} \mathcal{C}(\psi)
 \end{eqnarray}
 (with $\mathcal{C}$ being an arbitrary function of $\psi$) in the gauge \eqref{gauge1} is equivalent to some metric $f(R)$ gravity. The identification 
 \begin{eqnarray}\label{potential}
    2 e^{-\frac{2\sqrt{3}\psi}{3}}V_{f(R)}(\psi) \equiv \mathcal{C}(\psi)
 \end{eqnarray}
 will provide us with a direct relationship between the $f$ function and the scalar field $\psi$. This is of course a consequence of the fact that we neglected the matter part of the action, so that by means of conformal transformations and redefinition of the initial scalar field, we can establish equivalence of any STT with some $f(R)$ theory. It is the addition of matter that severely limits the number of equivalent models. 
 
 As an example, let us consider the following Lagrangian in \eqref{gauge1} gauge:
 \begin{equation}\label{lag2}
\mathcal{L} = \frac{1}{2}(-\dot{\xx}^2+\dot{\psi}^2) - 2 e^{\sqrt{3}(\lambda_1 \xx - \lambda_2\psi)}
\end{equation}
where $\lambda_1, \lambda_2 \in \mathbb{R}$. From the comparison with \eqref{generalLagrangian}, it is clear that $\lambda_1 = 1$ must hold for the equivalence to be present. Hamiltonian is easily constructed:
\begin{equation}\label{ham}
\mathcal{H} = \frac{1}{2}(- p^2_\xx+p^2_\psi ) +2 e^{\sqrt{3}(\lambda_1 \xx - \lambda_2\psi)}.
\end{equation}
This Hamiltonian admits a first order integral of motion of the form \cite{sorin}:
\begin{equation}
I = \alpha p_\psi + \beta p_\xx ,
\end{equation}
with $\alpha, \beta  \in \mathbb{R}$ if $\alpha=\frac{\lambda_2}{\lambda_1} \beta\equiv \lambda_2\beta$. 
Now, if we want to reproduce the original scalar-tensor theory from which the Hamiltonian (\ref{ham}) originates, we must compare the functions of corresponding separated variables in (\ref{potential}). For a general case, one cannot determine the functions $\cv$ and $\ca$ uniquely using the condition:
\begin{equation}
\frac{\cv(\psi)}{\ca^2(\psi)} = e^{-\sqrt{3}\lambda_2 \psi},
\end{equation}
but for $f(R)$ gravity, it becomes possible to obtain direct relation between the $\lambda_2$ and a particular choice of the $f$ function. To achieve this goal, we need to remember that the $\ca$ function for $f(R)$ theories is not arbitrary, but equal to:
\begin{equation}
\ca(\psi) = \Phi(\psi) = e^{\frac{\sqrt{3}}{3}\psi},
\end{equation}
which leaves us with:
\begin{equation}
\cv(\psi(\Phi)) \propto e^{\frac{\sqrt{3}(2 - 3\lambda_2)}{3}\psi}=e^{\frac{\sqrt{3}(2 - 3\lambda_2)}{3}\sqrt{3}\ln \Phi}\propto \Phi^{2 - 3\lambda_2}.
\end{equation}
But we know that $R = \frac{d\cv}{d\Phi}$ and $\Phi = \frac{df}{dR}$, so:
\begin{equation}
R \propto \Phi^{1-3\lambda_2} \rightarrow \Phi \propto R^{\frac{1}{1-3\lambda_2}},
\end{equation}
which means that (cf. \cite{Barrow2018}):
\begin{equation}
f(R) = c_1 R^\frac{2 - 3\lambda_2}{1 - 3\lambda_2} + c_2.
\end{equation}
Therefore, the exponent of the curvature can take all possible values except for unity, excluding GR with cosmological constant (since it does not fulfill the requirement $d^2f/dR^2\neq0$). 

\subsection{Hybrid theories}\label{hybrid_th}

Let us consider the hybrid metric-Palatini theories of gravity described by the action:

\begin{equation}
    \begin{split}
        S[g_{\mu\nu}, \Gamma^\alpha_{\mu\nu}, \chi]&  = \frac{1}{2\kappa^2}\int_\Omega d^4x\sqrt{-g} \left(\Omega_A R(g) + f(\mathcal{R}(g, \Gamma)\right) \\
        &+ S_\text{matter}[g, \chi]\,.
    \end{split}
\end{equation}

Here, $\Omega_A$ is a parameter and $f$ function is a correction term depending on the curvature scalar constructed \`{a} la Palatini, i.e. being a function of the metric and the connection, treated now as an independent quantity. The connection turns out to be an auxiliary field and can be eliminated, but unlike in the Palatini $f(R)$ gravity, the scalar field is dynamical. Therefore, hybrid theory is in fact a metric one, and has the following scalar-tensor representation:

\begin{equation}\label{hybridaction}
\begin{split}
   S[g_{\mu\nu} & , \Phi, \chi] =  \frac{1}{2\kappa^2}\int_\Omega d^4x \sqrt{-g} \Big[(\Omega_A + \Phi)R(g) \\
   &+ \frac{3}{2\Phi}g^{\mu\nu}\partial_\mu\Phi\partial_\nu\Phi - V_\text{hybrid}(\Phi)\Big]+ S_\text{matter}[g, \chi]. 
\end{split}
\end{equation}

We can immediately identify the four functions of the scalar field:
$$\ca(\Phi) = \Omega_A + \Phi,\quad\cb(\Phi) = -\frac{3}{2\Phi},$$$$\quad\cv(\Phi) = V_\text{hybrid}(\Phi),\quad\alpha(\Phi) = 0.$$

For this set of functions,  we choose $\Phi > 0$ to avoid dealing with phantom field and, consequently, $\text{sign}(\mathcal{I}) = -1$ (see section \ref{einsteinframes}). Upon integrating, the invariant scalar field will have the following form:
\begin{eqnarray}
   \psi = \pm 2\sqrt{3}\arctan\sqrt{\frac{\Phi}{\Omega_A}}
\end{eqnarray}
which, after inverting the relation, will allow us to express the initial scalar field in terms of the invariant one:
\begin{eqnarray}
   \Phi = \Omega_A \tan^2\left(\frac{\psi}{2\sqrt{3}}\right).
\end{eqnarray}
Let us now consider the Starobinsky-like quadratic correction of the Palatini form, i.e. $f(\mathcal{R})=\beta \mathcal{R}^2$. For this theory, the invariant potential can be also written as a function of the invariant scalar field:
\begin{eqnarray}
   \tilde{V}(\psi) = \frac{\cv(\Phi(\psi))}{\ca(\Phi(\psi))} = \frac{1}{4\beta}\sin^4\left(\frac{\psi}{2\sqrt{3}}\right).
\end{eqnarray}
Therefore, the MSS Lagrangian density, in the gauge $N = \ca^{-\frac{1}{2}}(\psi)e^{ \frac{\sqrt{3}}{2} \xx}$, will be:
\begin{eqnarray}
   \mathcal{L}_\text{MSS} = -\frac{1}{2}(\dot{\xx}^2 + \dot{\psi}^2) - \frac{1}{4\beta}e^{\sqrt{3} \xx}\sin^4\left(\frac{\psi}{2\sqrt{3}}\right)\,.
\end{eqnarray}
We perform one more coordinate change:
\begin{eqnarray}
   \xx & = &\frac{1}{\sqrt{3}}(\hat{\xx}- \ln 3),\label{par1}\\
   \psi & = &\frac{i}{\sqrt{3}}\hat{\psi}\label{par2}
\end{eqnarray}
so that the MSS Lagrangian density will get a multiplicative factor of $\frac{1}{3}$, which will be omitted in the following calculations, and the sine function will be changed to its hyperbolic counterpart:
\begin{eqnarray}\label{hybrid_lagr}
   \mathcal{L}_\text{MSS} = \frac{1}{2}(-\dot{\hat{\xx}}^2 + \dot{\hat{\psi}}^2) - \frac{1}{4\beta}e^{\pm\hat{\xx}}\sinh^4\left(\frac{\hat{\psi}}{6}\right)\,.
\end{eqnarray}
This Lagrangian is a special case of a more general family of Lagrangians \cite{sorin}:
\begin{equation}\label{sorin_lagr}
\mathcal{L} =  \frac{1}{2}(-\dot{x}^2 + \dot{y}^2) - c\: e^{x} \sinh^{\frac{-2\mu - 1}{\mu}}(\mu y),
\end{equation}
where $c, \mu$ are real numbers, possessing first-order integral of motion of the form:
\begin{eqnarray}\label{intofmot}
   I = e^{\mu x}\left[\cosh{(\mu y)}\: p_{x} + \sinh{(\mu y)}\:p_{y}\right].
\end{eqnarray}
The hybrid theories in the parametrization \eqref{par1}-\eqref{par2} correspond to $\mu = -\frac{1}{6}$ and $c = \frac{1}{4\beta}$, as it can be easily seen by comparing \eqref{sorin_lagr} and \eqref{hybrid_lagr}. In this case, the integral of motion takes the form \footnote{Although the momentum $p_{\hat{\psi}}$ is imaginary, its composition with the hyperbolic sine, having as an argument an imaginary quantity $\hat{\psi}$, will produce a real quantity. Therefore, the change of coordinates \eqref{par1}-\eqref{par2} does not entail any physical consequences and can be viewed as a mathematical trick only.}:
\begin{eqnarray}\label{intofmot}
   I = e^{-\frac{1}{6}\hat{\xx}}\Big[\cosh\left(\frac{\hat{\psi}}{6}\right) \:p_{\hat{\xx}} - \sinh\left(\frac{\hat{\psi}}{6} \right)\:p_{\hat{\psi}}\Big]\,.
\end{eqnarray}

It can be easily shown that the quadratic correction is the only one of the form $f(\mathcal{R}) = \beta_n \mathcal{R}^n$ leading to the integral of motion of the form \eqref{intofmot}.

\subsubsection*{Example: choice of initial condition in hybrid gravity}

As an example illustrating how the choice of present-day values of scalar field affects the late-time expansion of the universe, let us consider the following model, which can be obtained from a hybrid $f(R)$ gravity:
\begin{equation}
    \begin{split}
        S[g_{\mu\nu}, \Phi, \chi] &  = \frac{1}{2\kappa^2}\int_\Omega d^4x \sqrt{-g} \Big[(\Phi + \Omega_A)R + \frac{3}{2\Phi}(\partial\Phi)^2\\
        & - \frac{\sigma}{\lambda + \Phi}\Big]  + S_\text{matter}[g_{\mu\nu}, \chi],
    \end{split}
\end{equation}
where $\sigma, \lambda$ are real numbers. In order to obtain realistic solutions, one needs to use the Friedmann constraint equation \eqref{friedmann}, with an additional assumption that $H_0 = 1$ at present, which amounts to rescaling the cosmic time in an appropriate way (by choosing the lapse function to be $N = H_0$). Then, the equation is a simple algebraic equation for $\Phi_0$, relating the current value of the field to the remaining parameters and the 'real' matter content $\Omega_{0,m}$, which is related to the observed one $\tilde{\Omega}_{0,m}$ via:
\begin{eqnarray}
\tilde{\Omega}_{0,m} = \frac{\Omega_{0,m}}{2\mathcal{A}(\Phi_0)}.
\end{eqnarray}
If $\Omega_{0,m} = 0.05$, then there is only baryonic luminous matter present in the observable universe; any other number between 0.05 and 0.3 denotes a mixture of the normal matter and the dark component. 

We present various possibilities for $\Omega_A=1$, $\sigma = 52$ and $\lambda=10$ in the diagrams below. In order to obtain the solutions, we used the Friedmann equation \eqref{friedmann}, together with equations \eqref{ev1} and \eqref{ev2}. We note that the unit time is simply $1/H_0$ in the rescaled units. The present moment corresponds to $t=0$.
\begin{figure*}[h]
    \centering
    \includegraphics[width=0.9\textwidth]{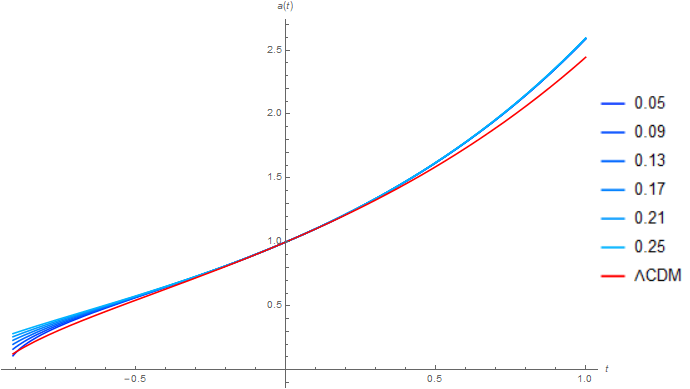}
    \caption{Scale factor as a function of dimensionless cosmic time for $\Omega_A=1$, $\sigma = 52$ and $\lambda=10$. Different colors represent different amounts of 'true' matter content $\Omega_{0,m}$. The red curve represents $\Lambda$CDM model.}
    \label{scale_f}
\end{figure*}

\begin{figure*}[h]
    \centering
    \includegraphics[width=0.8\textwidth]{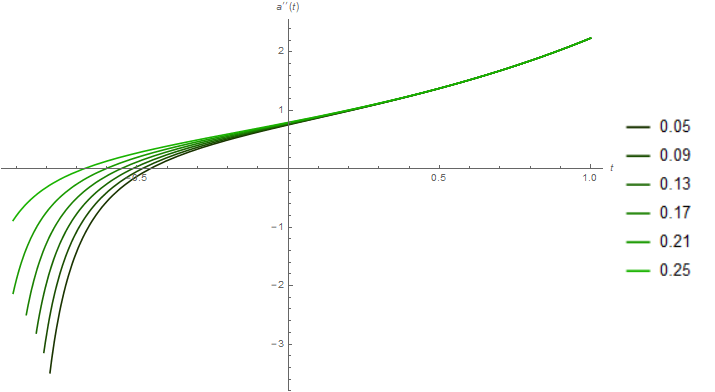}
    \caption{Second derivative of the scale factor as a function of dimensionless cosmic time for $\Omega_A=1$, $\sigma = 52$ and $\lambda=10$. Different colors represent different amounts of 'true' matter content.}
    \label{sec_der}
\end{figure*}

\begin{figure*}[h]
    \centering
    \includegraphics[width=0.8\textwidth]{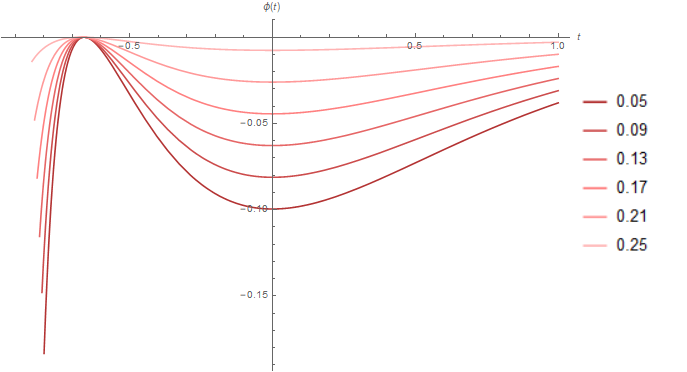}
    \caption{Scalar field as a function of dimensionless cosmic time for $\Omega_A=1$, $\sigma = 52$ and $\lambda=10$. Different colors represent different amounts of 'true' matter content.}
    \label{sf}
\end{figure*}

Figure (\ref{scale_f}) represents the evolution of the scale factor. We see that the more dark matter in the universe, the less steep the slope of the curve for negative values of the cosmic time (corresponding to moments preceding the present), therefore - the older the universe. Also, in the model we present, the universe is going to expand slightly faster than predicted by the $\Lambda$CDM model in the near future. As far as the second derivative of the field is concerned, shown in Fig. (\ref{sec_der}), its zero marks the moment when the universe stopped being matter-dominated (and thus the expansion decelerated) and started being dominated by dark energy. Its role is played here by the potential of the scalar field, which converges to a constant value for a large time. The scalar field, presented in Fig. (\ref{sf}) takes only negative values and approaches zero at larger times. 


\subsection{A comment on Einstein frame representation of different theories}\label{einsteinframes}
It turns out that, when matter is not taken into account, any ST theory becomes equivalent to some $f(R)$ gravity.
Any (non-degenerate) STT in the Einstein frame has $\bar\cb={1\over 2}$ or $\bar\cb=-{1\over 2}$. The first case corresponds to metric $f(R)$ theory and to a sub-class of hybrid metric-Palatini gravity. Following our earlier paper \cite{borowiec2020}, the
second case corresponds to a different sub-class of hybrid metric-Palatini formulation (discussed below). 
Finally, in degenerate case $\bar\cb=0$ not studied in this paper,  one can find the relation with purely Palatini $f(R)$-theory. This corollary follows directly from analysis of the sign of $\mathcal{I}$:
\begin{eqnarray}
   \text{sign}(\mathcal{I}) = -\text{sign}(m)=\begin{cases}
       +1& \text{ metric and hybrid}, \\
       0 &\text{ Palatini}, \\
       -1& \text{ hybrid}, 
   \end{cases}
\end{eqnarray}
which is related to the Einstein frame kinetic coupling through \eqref{transf2} and \eqref{invfield}. The quantity $\mathcal{I}$ cannot change the sign under conformal transformation and/or redefinition of the scalar field, since it transforms in the following way:
\begin{eqnarray}
   \bar{\mathcal{I}}(\bar{\Phi}) = e^{2\,\check{\gamma}(\bar{\Phi})}\Big(\frac{d\Phi}{d\bar{\Phi}}\Big)^2\mathcal{I}(\check{f}(\bar{\Phi})).
\end{eqnarray}
In case of the metric $f(R)$ gravity, one can always choose the Einstein frame with $\bar{B}=\frac{\text{sign}(\mathcal{I})}{2} =\frac{1}{2}$ by performing the following transformation:
\begin{subequations}
\begin{align}
    \bar{\Phi}(\Phi) &= \pm\sqrt{3}\ln\frac{\Phi}{\Phi_0}, \\
    \check{\gamma}(\bar{\Phi}) &= \mp\frac{\bar{\Phi}}{2\sqrt{3}}-\frac{1}{2}\ln \Phi_0.
\end{align}
\end{subequations}
On the other hand, hybrid formalism introduces two possible ways of choosing a self-consistent transformation for the scalar field, depending on its sign. To get the Einstein frame from \eqref{hybridaction}, one needs to perform a conformal transformation defined by:
\begin{eqnarray}
   \check{\gamma}(\bar{\Phi}) = -\frac{1}{2}\ln (\Omega_A + \Phi(\bar{\Phi})),
\end{eqnarray}
where the dependence between the new and old field needs to be determined using \eqref{transf2}. One gets:
\begin{eqnarray}
   \pm \frac{1}{2} = \left(\frac{d\Phi}{d\bar{\Phi}}\right)^2\frac{ 3\Omega_A}{2\Phi(\Omega_A + \Phi)^2},
\end{eqnarray}
where minus sign is chosen when $-\Omega_A < \Phi < 0$, and the plus sign if $\Phi > 0$\footnote{We neglect the case when $\Phi < -\Omega_A$, as it would lead to a negative effective gravitational constant.}, resulting in \cite{bronnikov}:
\begin{eqnarray}
\bar{\Phi}(\Phi) = 
   \begin{cases}
       \pm 2\sqrt{3\:}\text{arctanh}\sqrt{\frac{-\Phi}{\Omega_A}},\quad -\Omega_A < \Phi < 0,\\
       \pm 2\sqrt{3}\:\text{arctan}\sqrt{\frac{\Phi}{\Omega_A}}, \quad\:\:\:\Phi > 0.
   \end{cases}
\end{eqnarray}
We note that the possibility of $\Phi$ crossing zero is not analyzed here. It is however possible that dynamically, for certain classes of scalar-tensor theories, the effective gravitational constant becomes negative \cite{ayuso} when an appropriate choice of initial conditions is made.

\section{Conclusions and perspectives}
In this paper, we revisited MSS formalism and presented a methodology allowing one to analyze any scalar-tensor theory in such a way, that the resulting 2-dimensional MSS metric has a vanishing curvature allowing for introducing flat (Cartesian) coordinates. This was achieved by exercising the freedom of choosing the lapse function, which played the role of a conformal factor on the MSS after an appropriate change of variables. Such a choice amounts to performing a conformal transformation of the (4-dimensional) spacetime metric, bringing any theory defined by initially unspecified set of four functions of the scalar field to the so-called invariant Einstein frame. In the Einstein frame, the kinetic terms for the metric and the scalar field do not mix, so the Cauchy problem becomes easier to formulate. Also, the Einstein frame seems ‘natural’ since it introduces conformal invariants as dynamic variables. Such invariants preserve their functional form under Weyl rescaling of the metric tensor and redefinition of the scalar field, therefore they can be used to classify mathematically equivalent theories. Using the Einstein frame comes at a price – Einstein frame, due to the presence of anomalous coupling between the matter and scalar field, leads to violation of the Weak Equivalence Principle. Therefore, in the paper we did not consider the Einstein frame to be physical, and warned against giving physical meaning to transformed variables, in which the equations appear easier. 

A considerable amount of attention was dedicated to the lapse function. It was shown that some class of singular mechanical systems with a lapse function can be equivalently represented by the constrained conservative mechanical systems which incorporate the lapse function into both MSS metric and potential. Also, the lapse function can be changed in an arbitrary way (since the theory is invariant under spacetime diffeomorphisms), providing more convenient parametrization when investigation behavior of the Universe at various energy scales. When an implicit dependence on the time is assumed, one can use the lapse function to bring the MSS metric to the Lorentzian form. Particularly, this technique turns out to be useful in the Wheeler-DeWitt quantization allowing one to obtain a preferred operator ordering and elude the problem mentioned in \cite{Hawking}.

Any scalar tensor theory in the cosmological context can be analyzed using the mathematical formalism describing motion of a particle moving in a 4-dimensional phase space. Applying the Hamiltonian analysis, one can ask whether there exist integrals of motion for a given potential. Existence of such integrals constrains the motion of the particle, since it imposes additional conditions the particle must fulfill. Therefore, since the number of solutions becomes reduced, the system of equations is easier to solve. For example, one can always use the zero-energy condition (Friedmann equation) to determine the value of one of the parameters characterizing motion of the particle, when the remaining three are given. This was shown in the last part of the paper, when based on the current value of the scale factor, Hubble parameter and rate of change of the scalar field, its possible value was determined (after specifying the amount of dark matter). It was shown that for some (exotic) potential, de Sitter phase emerged and the expansion became accelerated. The question of whether such potential produces viable inflationary parameters remains open. 

As far as the integrals of motion are concerned, choosing an appropriate gauge (the lapse function) allowed us to establish a class of $f(R)$ theories of gravity which posses an additional first-order integral, analogous to the total momentum of the particle. The same procedure was repeated for hybrid theories of gravity, and it turned out that the only theory admitting first-order integral of motion is the Starobinsky model. It must be noted, however, that the integrals of motion were found for coordinates that did not have a physical interpretation (one of them being imaginary). Therefore, after finding such solutions, one must transform back to the physical frame. The analysis of hybrid theories revealed an interesting problem of possibility of changing sign of the gravitational constant caused by the scalar field crossing the zero value. Such a change of sign would lead to change of MSS metric signature. The possibility of signature change for the spacetime metric has been discussed for a long time in the literature, see for example in \cite{sakharov}-\cite{bor07}, but the issue remains unexplored for hybrid models. 

\section*{Acknowledgments}
This work has been  supported by the Polish National Science Center (NCN), project UMO-2017/27/B/ST2/01902
and benefited from COST Action CA15117 (CANTATA), supported by COST
(European Cooperation in Science and Technology). 

\appendix
  
\section{MSS Christoffel symbols, Gauss curvature in different coordinate systems  }

We began with the  MSS metric \eqref{metric2d} written down in canonical coordinates
that are natural coordinates   $(a, \Phi)$ inherited from the four-dimensional  STT in a specific frame.  Not all functions determining a frame enter the metric.  
Christoffel symbols can be decomposed, for convenience, into $a$  and $\Phi$ dependent parts:
\begin{subequations}
\begin{align}
\begin{split}
    & G^1_{\:11} = \frac{\ca(\Phi)\cb(\Phi)+3(\ca'(\Phi))^2}{a\left(2\ca(\Phi)\cb(\Phi)+ 3(\ca'(\Phi))^2\right)}\\
    &\quad\quad\doteq a^{-1} \gamma^1_{\:11}(\Phi)\,, 
\end{split}\\
\begin{split}
    G^1_{\:12} =\: & G^1_{\:21} = -\frac{\ca'(\Phi)\cb(\Phi)}{4\ca(\Phi)\cb(\Phi)+ 6(\ca'(\Phi))^2}\doteq  \gamma^1_{\:12}(\Phi)\,,
\end{split} \\
\begin{split}
    & G^1_{\:22} = \frac{a\Big(\cb(\Phi)^2-\cb'(\Phi)\ca'(\Phi)+2\cb(\Phi)\ca''(\Phi)\Big)}{4\ca(\Phi)\cb(\Phi)+ 6(\ca'(\Phi))^2} \\
    & \quad\quad\doteq a^{} \gamma^1_{\:22}(\Phi)\,,
\end{split} \\
\begin{split}
    & G^2_{\:11} = -\frac{3\ca(\Phi)\ca'(\Phi)}{a^2\left(2\ca(\Phi)\cb(\Phi)+ 3(\ca'(\Phi))^2\right)}\\
    &\quad\quad\doteq a^{-2} \gamma^2_{\:11}(\Phi)\,,
\end{split} \\
\begin{split}
    G^2_{\:12} =\: & G^2_{\:21} = \frac{3\ca(\Phi)\cb(\Phi)+3(\ca'(\Phi))^2}{a\left(2\ca(\Phi)\cb(\Phi)+ 3(\ca'(\Phi))^2\right)}\\
    &\quad\quad\doteq a^{-1} \gamma^2_{\:12}(\Phi)\,,
\end{split} \\
\begin{split}
    & G^2_{\:22} = \frac{3\ca'(\Phi)\cb(\Phi)+2\ca(\Phi)\cb'(\Phi)+6\ca'(\Phi)\ca''(\Phi)}{4\ca(\Phi)\cb(\Phi)+ 6(\ca'(\Phi))^2}\\
    &\quad\quad\doteq  \gamma^2_{\:22}(\Phi)\,.
\end{split}
\end{align}
\end{subequations}
Since in two-dimensions Einstein tensor vanishes, Ricci tensor takes the form $R^\texttt{MSS}_{ij} =\frac{1}{2} m_{ij} \,R^\texttt{MSS}$, where the Ricci scalar (double of Gauss curvature):
\begin{equation}
R^\texttt{MSS}(a,\Phi) = \frac{-\left(\ca \ca' \cb\right)'+3\ca  \ca'' \cb}{2\,a^3\left(2\ca \cb+3(\ca')^2\right)}
\end{equation}
determines Riemann curvature tensor: $R^\texttt{MSS}_{1212}= 1/2\,R^\texttt{MSS}\,(m_{11}m_{22}-m^2_{12})$. It implies that $R^\texttt{MSS}$ is singular at $a=0$ or when $\ci(\Phi)=0$, i.e. at $\Phi_s$. For checking MSS flatness condition, i.e. $R^\texttt{MSS}\equiv 0$, we switch to the  universal isothermal  coordinates setting the metric \eqref{metric2d} into conformally flat form \eqref{mss-2} with two possibilities for the signature Euclidean or Lorentzian ones. The last depends on the signature of the original metric \eqref{metric2d}.

In this coordinates, the  connection coefficients take very simple, yet general form  \eqref{christfr})
\begin{subequations}
	\begin{align}
		& G^2_{\:11}\: =\text{sign}(m)\frac{\ca'(\psi )}{4 \ca(\psi )}\,,\\
		& G^2_{\:22}\: =\:	G^1_{\:12} =\: G^1_{\:21}=\:-\frac{\ca'(\psi )}{4 \ca(\psi )}  \,, \\
		&G^2_{\:12}\: =\: G^2_{\:21}\: =\:G^1_{\:11}\: =\frac{\sqrt{3}}{4}\,,\quad G^1_{\:22}\:=\:-\text{sign}(m)\frac{\sqrt{3}}{4} \,.  
	\end{align}
\end{subequations}
MSS Ricci scalar  
\begin{equation}\label{gauss}
	R^\texttt{MSS}(\aa,\xx)=-\text{sign}(m)\,e^{-\frac{\sqrt{3}\xx}{2}} \frac{\left(-\ca'(\psi )^2+\ca(\psi ) \ca''(\psi)\right)}{2 \ca(\psi)^{3/2}}
\end{equation}
cannot take constant value unless $R_{MSS}=0$ which is possible only for $\ca(\psi)\sim\exp{(\sigma\psi)}$, where  $\sigma\in\mathbb{R}$
\footnote{Equivalently, the original $\ca (\Phi)\sim \exp{\left(\sigma\int \frac{\sqrt{\pm\,(2\ca \cb+3( \ca')^2)}}{\ca}d\Phi\right)}$.}. In this case the metric is flat and, for $\mbox{sign}(m) = -1$, can be reduced further (locally) to the Lorentzian form
$du^2-dv^2$ by the following change of  coordinates (for $\sigma\neq \beta=\frac{\pm\sqrt 3}{4}$)

$$u=e^{( \sigma-\beta ) (\psi-\xx)}+e^{(\beta +\sigma ) (\psi+\xx)}\,,$$$$
v=-e^{(\sigma-\beta ) (\psi-\xx)}+e^{(\beta +\sigma ) (\psi+\xx)}\,,$$ 
then $e^{2 \sigma  \psi +2 \beta  \xx}  \left(- d\xx^2 + d\psi^2 \right) = \frac{1}{4\left(\sigma ^2-\beta ^2\right)}(-dv^2 +du^2)$.

 For $\sigma=\beta=\pm\frac{\sqrt 3}{4}$ one takes
 $$u=\pm\frac{1}{\sqrt 3} e^{\pm\frac{\sqrt 3}{2}(\psi+\xx)}+\frac{1}{2}(\psi-\xx)\, ,$$$$ v= \pm\frac{1}{\sqrt 3} e^{ \pm\frac{\sqrt 3}{2} (\psi+\xx)}-\frac{1}{2}(\psi-\xx)$$
so that: $$ e^{\pm\frac{\sqrt 3}{2}( \psi + \xx)}  \left(- d\xx^2 + d\psi^2 \right)= (-dv^2 + du^2)$$.

If $\mbox{sign}(m) = 1$, then the metric is Euclidean, but one can define $\psi = i \bar{\psi}$, and use the following change of coordinates:
$$u=e^{( i \sigma-\beta ) (\bar{\psi}-\xx)}+e^{(i \sigma + \beta) (\bar{\psi}+\xx)}\,,$$$$
v=-e^{(i \sigma-\beta ) (\bar{\psi}-\xx)}+e^{(i \sigma + \beta) (\bar{\psi}+\xx)}\,$$
 

As it was mentioned above, such coordinates, mixing the scale factor with the scalar field, are not interesting from the point of view of physical interpretation.

\end{document}